\documentclass[aps,prl,twocolumn,superscriptaddress]{revtex4} 

\usepackage [latin1] {inputenc}
\usepackage {t1enc}
\usepackage [dvips] {graphicx}
\usepackage {verbatim}
\usepackage {epsfig}
\usepackage {amsmath}
\usepackage {amssymb}

\newcommand{\eqnref}[1]{(\ref{#1})}		
\newcommand{\figref}[1]{Fig.~\ref{#1}}	

\newcommand{\kpm}{\ket{\alpha^{\scriptscriptstyle +}_{\scriptscriptstyle -}}}	
\newcommand{\kpp}{\ket{\alpha^{\scriptscriptstyle +}_{\scriptscriptstyle +}}}
\newcommand{\kmp}{\ket{\alpha^{\scriptscriptstyle -}_{\scriptscriptstyle +}}}

\newcommand{\vacr}{|vac\rangle}						
\newcommand{\bra}[1]{\langle #1|}					
\newcommand{\ket}[1]{|#1\rangle}					

\newcommand{\tmc}[2]{|\gamma_{#1}(#2)\rangle}		
\newcommand{\smc}[1]{|\xi(#1)\rangle}				

\newcommand{\usmc}[1]{|\xi'(#1)\rangle}					
\newcommand{\utmc}[2]{|\gamma_{#1}'(#2)\rangle}			
\newcommand{\ptmc}[2]{|\tilde{\gamma}_{#1}'(#2)\rangle}	

\newcommand{\utmcbraket}[2]{\langle\gamma_{#1}'(#2)|\gamma_{#1}'(#2)\rangle}	

\begin{document}

\title{A Hybrid Long-Distance Entanglement Distribution Protocol}

\date{\today}

\author{J. B. Brask}\affiliation{QUANTOP, The Niels Bohr Institute, University of Copenhagen, 2100 Copenhagen \O, Denmark}
\author{I. Rigas}\affiliation{Department of Optics, Facultad de F\'isica, Universidad Complutense Madrid, 28040 Madrid, Spain}
\author{E. S. Polzik}\affiliation{QUANTOP, The Niels Bohr Institute, University of Copenhagen, 2100 Copenhagen \O, Denmark}
\author{U. L. Andersen}\affiliation{Department of Physics, Technical University of Denmark, Building 309, 2800 Lyngby, Denmark}
\author{A. S. S\o rensen}\affiliation{QUANTOP, The Niels Bohr Institute, University of Copenhagen, 2100 Copenhagen \O, Denmark}

\begin{abstract}
We propose a hybrid (continuous-discrete variable) quantum repeater protocol for distribution of entanglement over long distances. Starting from entangled states created by means of single-photon detection, we show how entangled coherent state superpositions, also known as `Schr\"odinger cat states', can be generated by means of homodyne detection of light. We show that near-deterministic entanglement swapping with such states is possible using only linear optics and homodyne detectors, and we evaluate the performance of our protocol combining these elements.
\end{abstract}


\maketitle


Two regimes for optical quantum computation and communication are commonly identified \cite{kok2007,braunstein2005}. In the discrete variable regime, single photons are the carriers of information with qubits encoded e.g.~in their polarisation or frequency, and measurements are performed by single-photon detection (SPD). In the continuous variable regime, information is encoded in continuous degrees of freedom of the electromagnetic field such as the field quadratures, which are measured via homodyne detection. Both regimes have their advantages and drawbacks. On the one hand, single photons are good for heralding events across lossy channels since their discrete nature implies that no partial loss can take place. Successful detection of a photon at the channel output unambiguously identifies a successful event (in the absence of dark counts). This is useful for entanglement generation \cite{cabrillo1999,dlcz}. On the other hand, certain tasks, such as quantum teleportation \cite{braunstein1998}, can be accomplished unconditionally with linear optics in the continuous regime while not in the discrete one, and homodyne detection allows quadrature measurements with much higher efficiency than what can be achieved with SPD at present.

\begin{figure}
\includegraphics[width=.45\textwidth]{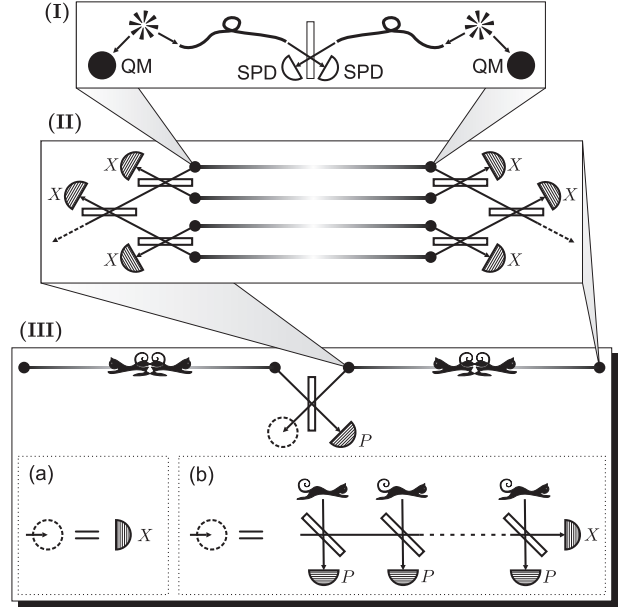}
\caption{\textbf{I} Generation of Bell-type entanglement between two nodes. Light from two-mode squeezing sources is mixed on a balanced beam splitter and a single photon is detected. The remaining modes are stored in quantum memories (QM). \textbf{II} Generation of approximate two-mode cat states. Pairs of Bell-like states are joined at balanced beam splitters, and the $X$-quadrature is measured at one output of each beam splitter. When the sum of the outcomes is close to zero, the state is kept, and the process is iterated. \textbf{III (a)} Simple entanglement swapping. Two two-mode cat states are joined on a balanced beam splitter, and the outputs are homodyned. Success is conditioned on an $X$-outcome close to zero. \textbf{III (b)} Improved entanglement swapping using $k$ auxilliary single-mode cat states of size (left to right) $2^{1/2}\alpha$, $2\alpha$,\ldots,$2^{k/2}\alpha$.}
\label{fig.felinerep}
\end{figure}

In quantum repeater schemes, entanglement distribution across a long lossy channel is achieved by combining entanglement generation over shorter segments with entanglement swapping and purification \cite{briegel1998}. Reaching high communication rates thus requires both a good generation scheme and efficient swapping. Repeaters combining continuous and discrete variables have recently been proposed. Ref.~\cite{vanLoock2006} relies on non-linear interactions of light with single spins in cavities, which is not easy to realise experimentally. Ref.~\cite{sangouard2009} is based on linear optics, but makes use of inefficient single-photon counters to perform entanglement swapping. Here we describe a quantum repeater (\figref{fig.felinerep}) which combines entanglement generation based on SPD with continuous variable techniques to efficiently distribute entanglement in the form of states resembling
\begin{equation}
\label{eq.tmcat}
\utmc{\theta}{\alpha} = e^{i\theta}\ket{\alpha}\ket{\alpha} + e^{-i\theta}\ket{-\alpha}\ket{-\alpha} ,
\end{equation}
where $\ket{\alpha}$ is a coherent state and $\theta$ is a phase (throughout this paper $\alpha$ is real). In quantum optics, coherent state superpositions of this form, and similarly for a single mode
\begin{equation}
\label{eq.smcat}
\usmc{\alpha} = \ket{\alpha} + \ket{-\alpha} ,
\end{equation}
are often called `Schr\"odinger cat' states. Such states are useful for a number of applications in quantum information, including fault-tolerant linear optical computation and quantum teleportation \cite{lund2008,phien2008}. Generating exact cat states is difficult, since it requires e.g.~very strong Kerr non-linearities \cite{glancy2008}. Approximate cat states can be generated by photon subtraction, as demonstrated recently by several groups \cite{ourjoumtsev2006,neergaardnielsen2006,takahashi2008,gerrits2009}, but unfortunately the average photon number is usually restricted to $\lesssim 1$.

We demonstrate that near-deterministic entanglement swapping of cat states using only linear optics and homodyning is possible, and we devise an efficient probabilistic scheme for generating states with many photons and very good overlap with exact squeezed single- and two-mode cat states. Our generation scheme is reminiscent of that of Ref.~\cite{ourjoumtsev}, where creation of squeezed single-mode cat states from photon-number states was demonstrated, but does not require input states with more than a single photon in each mode and takes advantage of quantum memories to significantly increase the rate. The elements of our repeater protocol are outlined in \figref{fig.felinerep}. In step (I), Fock state entanglement between two separated nodes is generated by means of SPD. In step (II), beam splitters and homodyne measurements are used to create two-mode cat states from the states generated in (I), and in step (III), the entangled cat states are swapped to larger distances. We discuss these steps one by one and then finally consider the protocol in its entirety.


As shown in \figref{fig.felinerep}, the entanglement generation step (I) can be implemented using two sources of two-mode squeezed states, realised using e.g.~parametric down conversion crystals or ensembles of $\Lambda$-type atoms \cite{simon2007,brask2008,dlcz}. One photonic mode from each source impinges on a central beam splitter, and a single photon is collected as indicated. Following a single SPD click, the remaining two modes are projected to a Bell-like state \cite{dlcz}
\begin{equation}
\label{eq.stokesbell}
\frac{1}{\sqrt{2}}(\ket{01} + \ket{10}) + O(\sqrt{p}) ,
\end{equation}
where the last term represents contributions from multiple excitations and is small when the pair production probability $p$ is small, i.e.~for weak squeezing. A setup of this type benefits from the discrete nature of single photons, with clicks heralding successful transmission.

In step (II), states of the form \eqnref{eq.stokesbell} are combined on balanced beam splitters, and the $X$-quadratures of two of the output modes are measured. The resulting state is kept, if the outcomes sum to a value within an interval $[-\Delta,\Delta]$. Otherwise it is discarded, and the process is restarted. Upon success, the protocol is iterated with the output states as new input states. To see that this scheme can produce cat-like states, it is illuminating to first consider the limit $\Delta \rightarrow 0$, $p \rightarrow 0$ and generation of a single-mode state (illustrated by, say, the left hand side of \figref{fig.felinerep} II only). We start from two sources, each producing a single excitation $\ket{1}$ corresponding to the standard harmonic oscillator wavefunction
\begin{equation}
\label{eq.singleex}
\psi_0(x) = \sqrt{2} \pi^{-1/4} e^{-\frac{1}{2} x^2} x .
\end{equation}
The joint wavefunction for both sources has the form $\psi_0(x)\psi_0(y)$. A balanced beam splitter (any non-diagonal unitary will do) is then applied to the pair of modes $x,y$, transforming the state to $\psi_0((x+y)/\sqrt{2})\psi_0((x-y)/\sqrt{2}) \propto e^{-(x^2+y^2)/2}(x^2-y^2) $, followed by a measurement of $y$. If we require $y = 0$, corresponding to $\Delta \rightarrow 0$, the resulting unnormalised output state takes the form $e^{-\frac{1}{2} x^2} x^2$. The process is now iterated, combining this state with the output from another pair of sources, etc. After $m$ iterations (in \figref{fig.felinerep} II, $m=2$), the final normalised output state wavefunction becomes
\begin{equation}
\label{eq.genfinalout}
\psi_m(x) = \Gamma(2^m+1/2)^{-1/2} e^{-\frac{1}{2} x^2} x^{2^m} .
\end{equation}
This expression is a symmetric, double-peaked function of $x$, which closely resembles the wavefunction of the state in Eqn.~\eqnref{eq.smcat}. In fact, defining $\smc{\alpha}$ to be the (normalised) single-mode cat state, we find that $\ket{\psi_m}$ is well approximated by $\hat{S}(2) \smc{\mu_m}$, where $\mu_m = \sqrt{2^{m}+1/2}$ and $\hat{S}(s)$ denotes squeezing in the variance of $X$ by a factor of $s$. The output state is thus very nearly a squeezed cat state. The fidelity $|\bra{\psi_m}\hat{S}(2)\smc{\mu_m}|^2$ of the actual output with respect to the squeezed cat exceeds 99\% for $m\geq 2$. Being squeezed does not render our states less useful. They are squeezed in amplitude, making them more robust against decoherence \cite{serafini2004}, and if desired they can be unsqueezed by local operations. The required squeezing of about 3dB or less is easily accessible in experiment. If the squeezing is introduced into the input Fock states, the single-mode setup corresponds to amplification of small cat states via homodyne detection \cite{lund2004,takeoka2007}. The scheme can be generalised from a single to an arbitrary number of modes by taking the mode of the single excitations from the sources to be a superposition of some other set of modes. Eq.~\eqnref{eq.stokesbell} for example corresponds to an excitation in a symmetric superposition of two spatial modes. More generally the new mode variables form a vector $\mathbf{x}$, and the source mode is given by $\mathbf{a}^\dagger \mathbf{x}$ where $\mathbf{a}$ is a unit vector. For a pair of sources, the single-mode scheme is applied to each pair of corresponding modes and we require $|\mathbf{a}^\dagger\mathbf{y}| \leq \Delta$, where $\mathbf{y}$ is the vector of measurement outcomes. Separating out the mode determined by $\mathbf{a}$, the final state takes the form $\ket{\psi_m}_\mathbf{a}\vacr$, where $\vacr$ is the vacuum state of the remaining modes.

\begin{figure}
\includegraphics[width=.4\textwidth]{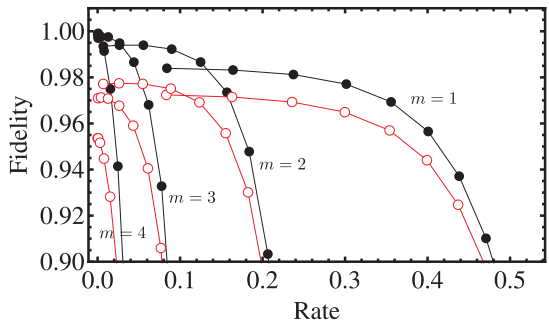}
\caption{(Colour online) Probabilistic generation of approximate cat states. The fidelity-rate trade-off is plotted for perfect input states (dots) and input states with a 1\% two-photon contribution (circles). The rate is measured in units of the source repetition rate.}
\label{fig.appcatgensimul}
\end{figure}

Step (II) may be implemented entirely on traveling light beams (see  \figref{fig.felinerep}), in which case simultaneous success for all measurements is required. The rate can, however, be significantly increased if the quantum states generated at each level of the protocol can be stored in memories, since in this case there is no requirement for simultaneity. The required homodyne measurements can be performed on retrieved light fields or via quantum nondemolition measurements directly on the memories \cite{appel2009}. To get a non-zero probability for successful generation one needs to take a non-zero value of $\Delta$, and there will be a trade-off between the rate of generation and the output state fidelity. To illustrate the trade-off with memories, we show in \figref{fig.appcatgensimul} a numerical simulation of a single-mode-cat generation. In the simulation, we include a double excitation contribution of 1\%, and we find that states with a fidelity of 90\% with respect to $\hat{S}(2)\smc{\mu_3}$, corresponding to an unsqueezed cat amplitude $\alpha = 2.9$, can be generated at a rate of $\sim 0.08$ measured in units of the source repetition rate, i.e.~the rate at which \eqnref{eq.singleex} can be produced.


As illustrated in \figref{fig.felinerep} IIIa, once entanglement has been established within two neighbouring segments, they can be connected by locally mixing the  modes at the central node on a balanced beam splitter and measuring the $X$- and $P$-quadratures at the two output ports. We will first consider entanglement swapping with ideal cat states and subsequently with the actual states produced in step (II) of our scheme. We denote the normalised equivalent of $\utmc{\theta}{\alpha}$ in Eq. \eqnref{eq.tmcat} by $\tmc{\theta}{\alpha}$. Note that the amount of entanglement in $\tmc{\theta}{\alpha}$ varies with $\theta$, but for coherent state amplitudes $|\alpha|^2 \geq 2$ the variation can be safely neglected and $\tmc{\theta}{\alpha}$ contains one ebit of entanglement. We consider entanglement swapping of two copies of $\tmc{0}{\alpha}$. The idea of the swapping procedure is that several terms in the input state may lead to the same measurement outcome. Conditioning on this outcome thus projects the output modes onto a superposition state, which is entangled. For coherent state inputs $\ket{\alpha_1},\ket{\alpha_2}$, the balanced beam splitter outputs coherent states with amplitudes $(\alpha_1 \pm \alpha_2)/\sqrt{2}$. Since each input is $\ket{\pm\alpha}$, the possible outputs are $\ket{\pm\sqrt{2}\alpha}$ and $\ket{0}$. There are two ways to obtain the latter, and hence measuring $X$ and conditioning on an outcome close to zero results in an entangled output state. More formally, the state prior to swapping is $\tmc{0}{\alpha}\tmc{0}{\alpha}$. Introducing the shorthand $\kpp = \ket{\alpha}\ket{\alpha}$, $\kpm = \ket{\alpha}\ket{-\alpha}$ etc., the (unnormalised) state after a $P$-measurement with an outcome $p_0$ becomes
\begin{equation}
\label{eq.entswap1pmeas}
\kpm\ket{\sqrt{2}\alpha}_x + \kmp\ket{-\sqrt{2}\alpha}_x + \utmc{\theta_0}{\alpha}\ket{0}_x ,
\end{equation}
where we have labelled the mode for which $X$ is measured by $x$, and $\theta_0 = -2\alpha p_0$. If $\alpha$ is large enough for $\ket{0}$ and $\ket{\sqrt{2}\alpha}$ to be nearly orthogonal, we see from \eqnref{eq.entswap1pmeas} that an $X$-measurement with an outcome close to zero will project the remaining modes to a perfect two-mode cat state. Since $\utmcbraket{\theta_0}{\alpha} \approx 2$, the probability for this successful outcome is roughly 1/2.

Probabilistic entanglement swapping of two-mode cat states is thus easy to implement. As we now demonstrate, the swapping can be made nearly deterministic using auxiliary single-mode cat states as a resource. The setup is shown in \figref{fig.felinerep} IIIb. Additional beam splitters are inserted between the first beam splitter output and the $X$-measurement. At the $j$'th beam splitter an auxiliary cat state $\smc{2^{j/2}\alpha}$ is injected, and a $P$-measurement with outcome $p_j$ is performed in one output port. We can understand what happens by considering just one auxiliary state. By mixing the output from the first beam splitter with a single-mode cat of amplitude $\sqrt{2}\alpha$ we arrange that each of the `failure' outputs above can combine with one term of the cat state to yield $\ket{0}$. An $X$-outcome of zero then projects the output to an entangled state. At the same time, the `success' output above splits into $\ket{\pm\alpha}$, leading to two possible measurement outcomes each of which still produces the entangled state $\utmc{\theta_0}{\alpha}$. We have thus increased the number of $X$-outcomes which lead to successful swapping. Formally, after the two $P$-measurements the (unnormalised) state is
\begin{align}
\label{eq.entswap2pmeas}
\ptmc{\theta_1}{\alpha} \ket{0}_x & + \utmc{\theta_0}{\alpha} (e^{i\nu} \ket{\alpha}_x + e^{-i\nu} \ket{-\alpha}_x) \\
                         & + \kpm \ket{2\alpha}_x + \kmp \ket{-2\alpha}_x  , \nonumber
\end{align}
where $\theta_1 = -2^{3/2}\alpha p_1$, $\nu$ is an unimportant phase and $\ptmc{\theta_1}{\alpha} = e^{i\theta}\kpm + e^{-i\theta}\kmp$ equals $\utmc{\theta_1}{\alpha}$ up to a local phase shift. Assuming that $\ket{0}$ and $\ket{\alpha}$ are nearly orthogonal, $X$-outcomes originating in the $\ket{0}$ or $\ket{\pm\alpha}$ terms of \eqnref{eq.entswap2pmeas} all project the output to a two-mode cat state. Only the extremal outcomes lead to failed swapping. Counting terms, the success probability is seen to be $3/4$. Generalising this to $k$ auxiliary states, we find that the success probability scales as $1-2^{-k-1}$. Fixed distinguishability of the terms requires $\alpha$ to scale as $\alpha \sim 2^{k/2}$. The failure probability thus scales inversely with the mean photon number $\alpha^2$ in the two-mode cats and the square root of the mean photon number in the largest single-mode cat. This result demonstrates that, given sufficiently large cat-state resources, entanglement swapping using only linear optics and homodyne measurements can be performed with success probability arbitrarily close to one, i.e.~near-deterministically.


We now turn to the assembly of all three steps -- generation of entanglement in steps (I), (II) and entanglement swapping with the generated states in step (III) -- into a complete quantum repeater protocol. The goal of the protocol is to establish a useful entangled state across some channel of length $L$. To this end, the channel is divided into $2^n$ segments of a shorter length $L_0$. Entanglement is generated in each segment separately and they are then connected by entanglement swapping. We will consider only simple swapping (\figref{fig.felinerep} IIIa) with no auxiliary states. The benchmark for the performance of a repeater is the rate at which final, entangled states can be generated over a distance $L$ with a fixed fidelity with respect to some ideal target state. Just as in step (II), one must choose a finite acceptance range $[-\delta,\delta]$ for the $X$-measurements during swapping, and for fixed $L$ there will be a trade-off between repeater rate and fidelity through the parameters $\Delta,\delta,p$. The target state is approached as $\Delta,\delta,p \rightarrow 0$. In this limit, the entangled states before swapping are given by $\ket{\psi_m}_+\vacr_-$, where $\pm$ stand for the symmetric and antisymmetric combinations $a \pm b$ of the spatial modes $a$, $b$. Conditioning on zero in all $X$-measurements, we find that for sufficiently large $m$, up to two $P$-space displacements which can be trivially canceled, the wavefunction after $n$ connections takes the form $\hat{S}_+(\frac{4}{k_n})\hat{S}_-(\frac{k_n}{2}) \tmc{\varphi_n}{\mu_m/\sqrt{2k_n}}$. Here $k_n = 2\sqrt{2}\coth(2^n \mathrm{arccoth}(1/\sqrt{2}))$ while $\varphi_n$ depends on the $P$-measurement outcomes. Because $k_n$ converges fast towards $2\sqrt{2}$, this state is essentially equivalent to $\hat{S}_a(\sqrt{2})\hat{S}_b(\sqrt{2}) \tmc{\varphi_n}{\mu_m/2^{5/4}}$. Our target state is thus a squeezed, nonlocal cat state. We remark that the $1.5\,\mathrm{dB}$ squeezing in the final state is local and can be locally undone, even though the squeezed mode for the states generated in step II is the nonlocal, symmetric mode `$+$'.


\begin{figure}
\includegraphics[width=.4\textwidth]{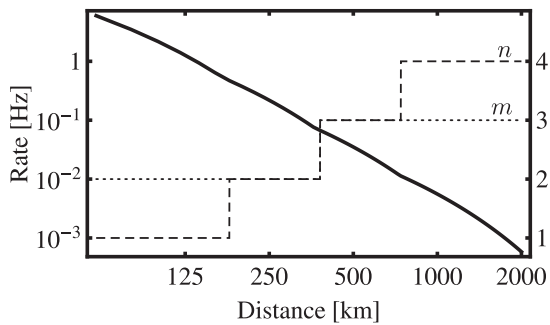}
\caption{Rate (solid line) of the full repeater protocol, including both generation and connection of entangled states. We assume a fixed final fidelity of 90\%, a fiber attenuation length of 20km and an SPD efficiency of 50\%. The rhs axis shows the optimal values of $n$ and $m$.}
\label{fig.repsimul}
\end{figure}

We numerically simulate the repeater protocol, optimising the rate as a function of distance for a fixed final fidelity of 90\% with respect to $\hat{S}_a(\sqrt{2})\hat{S}_b(\sqrt{2}) \tmc{-\varphi_n}{\mu_m/2^{5/4}}$. We optimise over the acceptance intervals $\Delta$ and $\delta$, the number of generation and connection steps $m$, $n$, and the pair production probability $p$. For runtime reasons we restrict the simulation to $m \leq 3$. We take into account losses during entanglement generation, assuming a fiber attenuation length of 20km and an SPD inefficiency of 50\%, and we take the source repetition time to be given by the classical communication time $L_0/c$, where $c$ is the speed of light. The result is shown in \figref{fig.repsimul}. At 1000km our protocol reaches a rate of 0.3 pairs/minute. The best proposed atomic-ensemble-based repeaters in the discrete variable regime achieve comparable rates. However to do so they require either high efficiency SPDs ($\gtrsim 90\%$), which unlike the efficient homodyne detectors employed here are not readily available in the lab, or more involved entanglement swapping procedures \cite{sangouard2008,brask2010}. In addition, for repeaters based on linear optics and SPD in the discrete variable regime, the success probability for entanglement swapping can never exceed $1/2$, whereas for the protocol presented here we have demonstrated that near-deterministic swapping and hence a much higher rate is possible when the swapping procedure of \figref{fig.felinerep} IIIa is replaced by that of IIIb. An analogous setup can also be used for near-deterministic teleportation. Thus the hybrid approach represents a promising avenue for reaching higher rates for entanglement distribution. Since our protocol uses only linear optics, light storage and retrieval, and single-photon and homodyne detectors, the means for a first experimental implementation are available in today's laboratories. An interesting variation of the present scheme might be obtained by integrating it with the proposal for generation of two-mode from single-mode cats of Ref.~\cite{sangouard2009}.


We acknowledge the EU FET-Open project COMPAS (212008), the FET-Proactive project Q-ESSENCE (248095), and the Danish National Research Foundation and Agency for Science, Technology, and Innovation.

\bibliography{Feline}

\end{document}